\documentclass[reprint,onecolumn,amsmath,amssymb,jap,aip]{revtex4-1}
\usepackage{graphicx}% Include figure files
\usepackage{dcolumn}% Align table columns on decimal point
\usepackage{bm}% bold math
\usepackage{epsfig}
\usepackage{mathptmx}
\usepackage{times}
\usepackage{amsmath}
\usepackage{amssymb}
\usepackage{dcolumn}
\usepackage{units}
\usepackage{upgreek}
\usepackage{tabularx}
\usepackage{todonotes}
\usepackage{threeparttable}
\usepackage[space]{grffile}
\usepackage[normalem]{ulem}
\newcommand{\paramdef}[1]{\textit{\lowercase{#1}}}
\newcommand{\vecdef}[1]{\textit{\uppercase{#1}}}
\newcommand{\scaldef}[1]{\textit{\lowercase{#1}}}
\newcommand{\keydef}[1]{`\textit{{#1}}'}
\DeclareMathOperator*{\argmaxA}{arg\,max} %
\usepackage{subfig}
\usepackage{multirow}
\usepackage{array}
\usepackage{xcolor}
\usepackage{soul}

\begin{document}
\title{In-memory hyperdimensional computing}
\author{Geethan Karunaratne} \affiliation{IBM Research -- Zurich, S\"{a}umerstrasse 4, 8803 R\"{u}schlikon, Switzerland.}\affiliation{Department of Information Technology and Electrical Engineering, ETH Z\"{u}rich, Gloriastrasse 35, 8092 Z\"{u}rich, Switzerland.}
\author{Manuel Le Gallo} \affiliation{IBM Research -- Zurich, S\"{a}umerstrasse 4, 8803 R\"{u}schlikon, Switzerland.}
\author{Giovanni Cherubini} \affiliation{IBM Research -- Zurich, S\"{a}umerstrasse 4, 8803 R\"{u}schlikon, Switzerland.}
\author{Luca Benini} \affiliation{Department of Information Technology and Electrical Engineering, ETH Z\"{u}rich, Gloriastrasse 35, 8092 Z\"{u}rich, Switzerland.}
\author{Abbas Rahimi} \email{abbas@ee.ethz.ch}\affiliation{Department of Information Technology and Electrical Engineering, ETH Z\"{u}rich, Gloriastrasse 35, 8092 Z\"{u}rich, Switzerland.}
\author{Abu Sebastian} \email{ase@zurich.ibm.com}\affiliation{IBM Research -- Zurich, S\"{a}umerstrasse 4, 8803 R\"{u}schlikon, Switzerland.}
\date{\today}
\begin{abstract}
Hyperdimensional computing (HDC) is an emerging computational framework that takes inspiration from attributes of neuronal circuits such as
hyperdimensionality, fully distributed holographic representation, and (pseudo)randomness. When employed for machine learning tasks such as learning
and classification, HDC involves manipulation and comparison of large patterns within memory. Moreover, a key attribute of HDC is its robustness to
the imperfections associated with the computational substrates on which it is implemented. It is therefore particularly amenable to emerging non-von
Neumann paradigms such as in-memory computing, where the physical attributes of nanoscale memristive devices are exploited to perform computation in
place. Here, we present a complete in-memory HDC system that achieves a near-optimum trade-off between design complexity and classification accuracy
based on three prototypical HDC related learning tasks, namely, language classification, news classification, and hand gesture recognition from
electromyography signals. Comparable accuracies to software implementations are demonstrated, experimentally, using 760,000 phase-change memory
devices performing analog in-memory computing.
\end{abstract}
\maketitle

\section{Introduction}
When designing biological computing systems, nature decided to trade accuracy for efficiency. Hence, one viable solution for continuous reduction in
energy is to adopt computational approaches that are inherently robust to uncertainty. Hyperdimensional computing (HDC) is recognized as one such
framework based on the observation that key aspects of human memory, perception, and cognition can be explained by the mathematical properties of
hyperdimensional spaces comprising high dimensional binary vectors known as hypervectors. Hypervectors are defined as $d$-dimensional, where
$\paramdef{d} \geq 1000$, (pseudo)random vectors with independent and identically distributed (i.i.d.) components~\cite{Kanerva98SDM}. When the
dimensionality is in the thousands, there exist a large number of quasiorthogonal hypervectors. This allows HDC to combine such hypervectors into new
hypervectors using well-defined vector space operations, defined such that the resulting hypervector is unique, and with the same dimension. A
powerful system of computing can be built on the rich algebra of hypervectors~\cite{Kanerva2009}. Groups, rings, and fields over hypervectors become
the underlying computing structures with permutations, mappings, and inverses as primitive computing operations.

In recent years, HDC has been employed for solving cognitive tasks such as Raven's progressive matrices \cite{Y2013emruliIJCNN} and
analogical reasoning \cite{Y2009slipchenkoITA}. HDC has also shown promising capabilities in machine learning applications that involve temporal
sequences such as text classification \cite{Y2000kanervaAMCSS}, biomedical signal processing \cite{Y2019rahimiProcIEEE,Y2019burrelloDATE}, multimodal
sensor fusion \cite{Y2015rasanenTNNLS}, and learning sensorimotor control for active perception in robots \cite{Y2019mitrokhinScienceRobotics}. The
training algorithm in HDC works in one or few shots, i.e., object categories are learned from one or few examples, and in a single pass over training
data as opposed to many iterations. In the aforementioned machine learning applications, HDC has achieved similar or higher accuracy with fewer
training examples compared to support vector machine (SVM) \cite{Y2019rahimiProcIEEE}, extreme gradient boosting \cite{Y2019changAICAS}, and
convolutional neural network (CNN) \cite{Y2019mitrokhinScienceRobotics}, and lower execution energy on embedded CPU/GPU compared to SVM
\cite{Y2018montagnaDAC}, CNN and long short-term memory \cite{Y2019burrelloDATE}.

HDC begins with representing symbols with i.i.d. hypervectors that are combined by nearly i.i.d.-preserving operations, namely
binding, bundling, and permutation, and then stored in associative memories to be recalled, matched, decomposed, or reasoned about. This chain
implies that failure in a component of a hypervector is not ``contagious" and forms a computational framework that is intrinsically robust to
defects, variations, and noise \cite{Kanerva2009}. The manipulation of large patterns stored in memory and the inherent robustness make HDC
particularly well suited for emerging computing paradigms such as in-memory computing or computational memory based on emerging nanoscale resistive
memory or memristive devices ~\cite{Y2013yangNatureNano,Y2017sebastianNatComm,Y2018zidanNatureElectronics,Y2018ielminiNatureElectronics}. In one such
work, 3D vertical resistive random access memory (ReRAM) device was used to perform individual operations for HDC~\cite{Y2016liIEDM,Y2017liVLSI}. In
another work, a carbon nanotube field effect transistor-based logic layer was integrated to ReRAMs, improving efficiency further~\cite{Y2018wuISSCC}.
However, these architectures resulted in limited application such as a single language recognition
task~\cite{Y2016liIEDM,Y2018wuISSCC}, or a restricted binary classification version of the same task~\cite{Y2018wuISSCC}; their evaluation is based
on simulations and compact models derived from small prototypes with only 256 ReRAM cells~\cite{Y2016liIEDM}, or a small 32-bit datapath for
hypervector manipulations that results in three orders of magnitude higher latency overhead~\cite{Y2018wuISSCC}.

In this paper, we propose a complete integrated in-memory HDC system in which all the operations of HDC are implemented on two planar memristive
crossbar engines together with peripheral digital CMOS circuits. We devise a novel way of performing hypervector binding entirely within a first
memristive crossbar using an in-memory read logic operation and hypervector bundling near the crossbar with CMOS logic. These key operations of HDC
co-operatively encode hypervectors with high precision, while eliminating the need to repeatedly program (write) the memristive devices.
In contrast, prior work on HDC using memristive devices did not employ in-memory logic operations for binding, instead a ReRAM-based
XOR lookup table \cite{Y2016liIEDM} or digital logic \cite{Y2018wuISSCC} were used. Moreover, the prior in-memory compute primitives for permutation
\cite{Y2016liIEDM} and bundling \cite{Y2018wuISSCC} resulted in repeated programming of the memristive devices which is prohibitive given the limited
cycling endurance. Associative memory search is performed in our architecture using a second memristive crossbar for in-memory dot product operations
on the encoded output hypervectors from the first crossbar, realizing the full HDC system functionality. The proposed combination of analog in-memory
computing with CMOS logic allows continual functioning of the memristive crossbars with desired accuracy for a wide range of multiclass
classification tasks. We verify the integrated inference functionality of the system through large-scale mixed hardware/software experiments in
which up to 49 $d=10,000$-dimensional hypervectors are encoded in $760,000$ hardware phase-change memory (PCM) devices performing analog in-memory
computing. Our experiments achieve comparable accuracies to the software baselines and surpass those reported in previous work on an emulated small
ReRAM crossbar \cite{Y2016liIEDM}. Furthermore, a complete system-level design of the in-memory HDC architecture synthesized using \unit[65]{nm} CMOS
technology demonstrates $>6\times$ end-to-end reductions in energy compared with a dedicated digital CMOS implementation. To
summarize, we map all operations of HDC either in-memory, or near-memory, and demonstrate their integrated functionality for three distinct
applications that are particularly well suited for HDC.

%-------------------------------------------------------------------------------------------------------------------------------------------------
% ///////////////////////////////////////////////// Concept of in-memory HD computing ////////////////////////////////////////////////////////////
%-------------------------------------------------------------------------------------------------------------------------------------------------
\section{The concept of in-memory HDC}\label{sec:inmemoryhd}

\begin{figure}[h!]
\centering
\includegraphics[width=\textwidth]{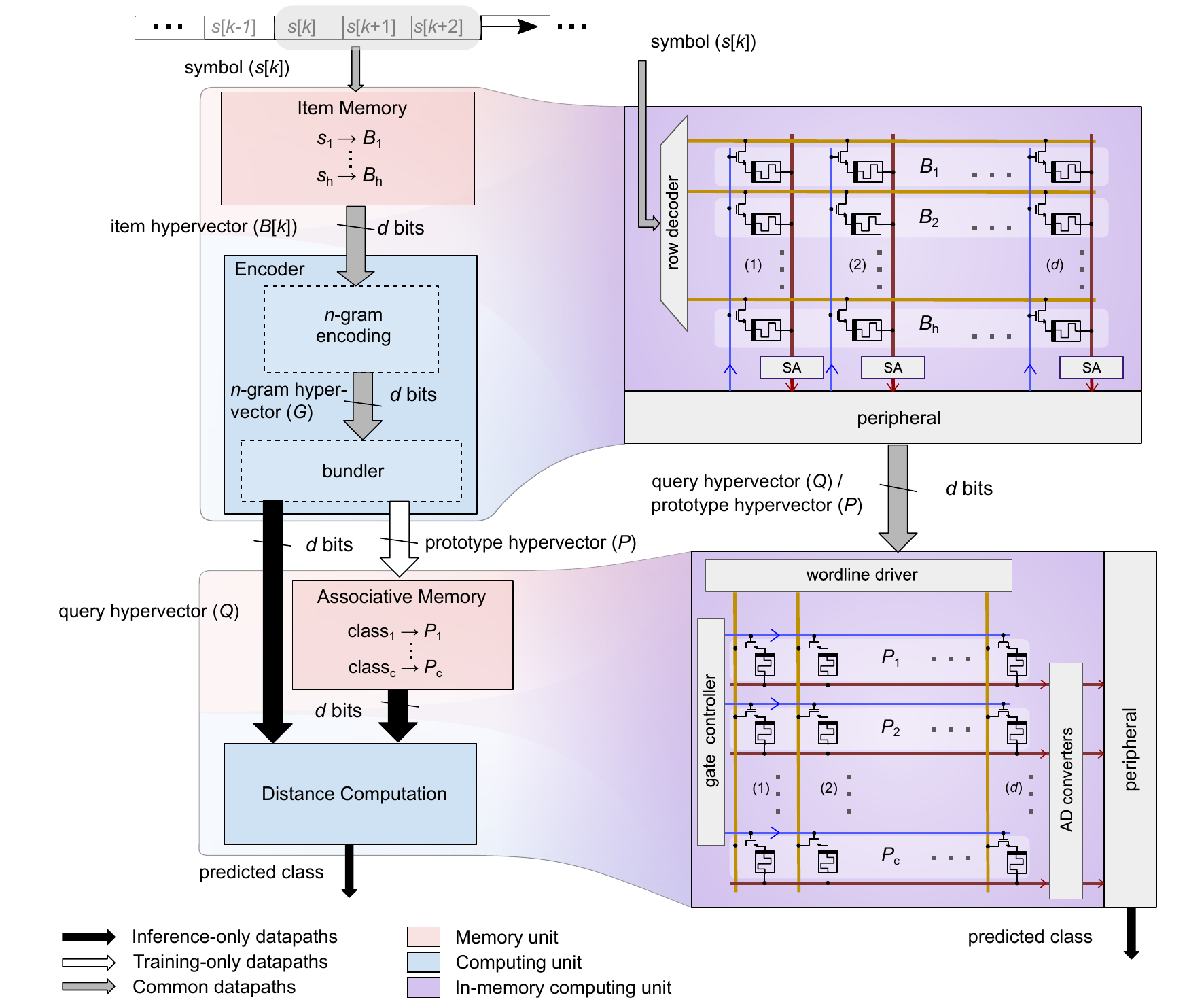}
\caption{\label{fig:1}\textbf{The concept of in-memory HDC} A schematic illustration of the concept of in-memory HDC shows the essential steps
associated with HDC (left) and how they are realized using in-memory computing (right). An item memory (IM) stores $h$, $d$-dimensional basis
hypervectors that correspond to the symbols associated with a classification problem. During learning, based on a labelled training dataset, an
encoder performs dimensionality preserving mathematical manipulations on the basis hypervectors to produce $c$, $d$-dimensional prototype
hypervectors that are stored in an associative memory (AM). During classification, the same encoder generates a query hypervector based on a test
example. Subsequently, an associative memory search is performed between the query hypervector and the hypervectors stored in the AM to determine the
class to which the test example belongs. In in-memory HDC, both the IM and AM are mapped onto crossbar arrays of memristive devices. The mathematical
operations associated with encoding and associative memory search are performed in-place by exploiting in-memory read logic and dot product
operations, respectively. A dimensionality of $d=10,000$ is used. SA: sense amplifier; AD converters: analog-to-digital converters.}
\end{figure}

When HDC is used for learning and classification, first, a set of i.i.d., hence quasiorthogonal hypervectors, referred to as basis hypervectors, are
selected to represent each symbol associated with a dataset. For example, if the task is to classify an unknown text into the corresponding language,
the symbols could be the letters of the alphabet. The basis hypervectors stay fixed throughout the computation. Assuming that there are $h$ symbols,
$\{s_i\}_1^h$, the set of the $h$, $d$-dimensional basis hypervectors $\{B_i\}_1^h$ is referred to as the item memory (IM) (see Fig. \ref{fig:1}).
Basis hypervectors serve as the basis from which further representations are made by applying a well-defined set of component-wise operations:
addition of binary hypervectors $[+]$ is defined as the component-wise majority, multiplication ($\oplus$) is defined as the component-wise
exclusive-OR (or equivalently as the component-wise exclusive-NOR), and finally permutation ($\rho$) is defined as a pseudo-random shuffling of the
coordinates. Applied on dense binary hypervectors where each component has equal probability of being zero or one~\cite{BSC96}, all these operations
produce a $\paramdef{d}$-bit hypervector resulting in a closed system.

Subsequently, during the learning phase, the basis hypervectors in the IM are combined with the component-wise operations inside an encoder to
compute for instance a quasiorthogonal $n$-gram hypervector representing an object of interest \cite{joshi2016}; and to add $n$-gram hypervectors from the same category of objects
to produce a prototype hypervector representing the entire class of category during learning. In the language example, the encoder would receive
input text associated with a known language and would generate a prototype hypervector corresponding to that language. In this case $n$ determines
the smallest number of symbols (letters in the example) that are combined while performing an $n$-gram encoding operation. The overall encoding
operation results in $c$, $d$-dimensional prototype hypervectors (referred to as associative memory (AM)) assuming there are $c$ classes. When the
encoder receives $n$ consecutive symbols, $\left\lbrace s[1], s[2], \hdots, s[n] \right\rbrace $, it produces an $n$-gram hypervector, $G$, given by:
\begin{equation}\label{eq:encoding}
G(s[1],s[2],\hdots,s[n]) =  B[1]\overline{\oplus} \rho(B[2]) \overline{\oplus} \hdots \overline{\oplus} \rho^{n-1}(B[n]),
\end{equation}
where $B[k]$ corresponds to the associated basis hypervector for symbol, $s[k]$. The operator $\overline{\oplus}$ denotes the exclusive-NOR, and
$\rho$ denotes a pseudo-random permutation operation, e.g., a circular shift by 1 bit. The encoder then bundles several such $n$-gram hypervectors
from the training data using component-wise addition followed by a binarization (majority function) to produce a prototype hypervector for the given
class.

When inference or classification is performed, a query hypervector (e.g. from a text of unknown language) is generated identical to the way the
prototype hypervectors are generated. Subsequently, the query hypervector is compared with the prototype hypervectors inside the AM to make the
appropriate classification. Equation~\ref{eq:am} defines how a query hypervector $Q$ is compared against each of the prototype hypervector $P_i$ out
of $c$ classes to find the predicted class with maximum similarity. This AM search operation can for example be performed by calculating the inverse
Hamming distance.
\begin{equation} \label{eq:am}
\mathrm{Class_{Pred}} = \argmaxA_{i\in \{1,...,c\}} \sum_{j=1}^d Q(j)\overline{\oplus}P_i(j)
\end{equation}

One key observation is that the two main operations presented above, namely, the encoding and AM search, are about manipulating and comparing large
patterns within the memory itself. Both IM and AM (after learning) represent permanent hypervectors stored in the memory. As a lookup operation,
different input symbols activate the corresponding stored patterns in the IM that are then combined inside or around memory with simple local
operations to produce another pattern for comparison in AM. These component-wise arithmetic operations on patterns allow a high degree of parallelism
as each hypervector component needs to communicate with only a local component or its immediate neighbors. This highly memory-centric aspect of HDC is
the key motivation for the in-memory computing implementation proposed in this work.

The essential idea of in-memory HDC is to store the components of both the IM and the AM as the conductance values of nanoscale memristive devices
\cite{Y2011chuaAPA,Y2015wongNatureNano} organized in crossbar arrays and enable HDC operations in or near to those devices (see Fig.~\ref{fig:1}).
The IM of $h$ rows and $d$ columns is stored in the first crossbar, where each basis hypervector is stored on a single row. To perform $\oplus$
operations between the basis hypervectors for the $n$-gram encoding, an in-memory read logic primitive is employed. Unlike the vast majority of
reported in-memory logic operations \cite{Y2010borghettiNature,Y2014kvatinskyTCAS}, the proposed in-memory read logic is non-stateful and this
obviates the need for very high write endurance for the memristive devices. Additional peripheral circuitry is used to implement the remaining
permutations and component-wise additions needed in the encoder. The AM of $c$ rows and $d$ columns is implemented in the second crossbar, where each
prototype hypervector is stored on a single row. During supervised learning, each prototype hypervector output from the first crossbar gets
programmed into a certain row of the AM based on the provided label. During inference, the query hypervector output from the first crossbar is input
as voltages on the wordline driver, to perform the AM search using an in-memory dot product primitive. Since every memristive device in the AM and IM
is reprogrammable, the representation of hypervectors is not hardcoded, as opposed to Refs. \onlinecite{Y2016liIEDM,Y2017liVLSI,Y2018wuISSCC} that
used device variability for projection.

This design ideally fits the memory-centric architecture of HDC, because it allows to perform the main computations on the IM and AM within the
memory units themselves with a high degree of parallelism. Furthermore, the IM and AM are only programmed once while training on a specific dataset.
and the two types of in-memory computations that are employed, involve just read operations. Therefore, non-volatile memristive devices are very well
suited for implementing the IM and AM, and only binary conductance states are required. In this work, we used PCM
technology~\cite{Y2010wongProcIEEE,Y2016burrJETCAS}, which operates by switching a phase-change material between amorphous (high resistivity) and
crystalline (low resistivity) phases to implement binary data storage (see Methods). PCM has also been successfully employed in novel computing
paradigms such as neuromorphic computing~\cite{Y2011kuzumNanoLetters,Y2016tumaNatNano,Y2018boybatNatComm,Y2018sebastianJAP} and computational
memory~\cite{Y2013wrightAFM,Y2017sebastianNatComm,Y2018legalloNatureElectronics,Y2018ielminiNatureElectronics}, which makes it a good candidate for
realizing the in-memory HDC system.

In the remaining part of the paper, we will elaborate the detailed designs of the associative memory, the encoder, and finally propose a complete
in-memory HDC system that achieves a near-optimum trade-off between design complexity and output accuracy. The functionality of the in-memory HDC
system will be validated through experiments using a prototype PCM chip fabricated in \unit[90]{nm} CMOS technology (see Methods), and a complete
system-level design implemented using \unit[65]{nm} CMOS technology will be presented.
%-------------------------------------------------------------------------------------------------------------------------------------------------
% ///////////////////////////////////////////////// ASSOCIATIVE MEMORY ////////////////////////////////////////////////////////////////////
%-------------------------------------------------------------------------------------------------------------------------------------------------
\section{The associative memory search module}\label{sec:am}

Classification involves an AM search between the prototype hypervectors and the query hypervector using a suitable similarity metric, such as the
inverse Hamming distance ($invHamm$) computed from Equation (\ref{eq:am}). Using associativity of addition operations, the expression in Equation
(\ref{eq:am}) can be decomposed into the addition of two dot product terms as shown in Equation~\ref{eqn:am2}
\begin{eqnarray}\label{eqn:am2}
\mathrm{Class_{Pred}} & = & \argmaxA_{i\in \{1,...,c\}} Q \cdot P_i + \overline{Q} \cdot \overline{P_i}\\\nonumber & \backsimeq &  \argmaxA_{i\in \{1,...,c\}} Q \cdot P_i
\end{eqnarray}
where $\overline{Q}$ denotes the logical complement of $Q$. Since the operations associated with HDC ensure that both the query and
prototype hypervectors have an almost equal number of zeros and ones, the dot product ($dotp$) $\argmaxA_{i\in \{1,...,c\}} Q \cdot P_i$
can also serve as a viable similarity metric.

\begin{figure}[h!]
\includegraphics[width=\textwidth]{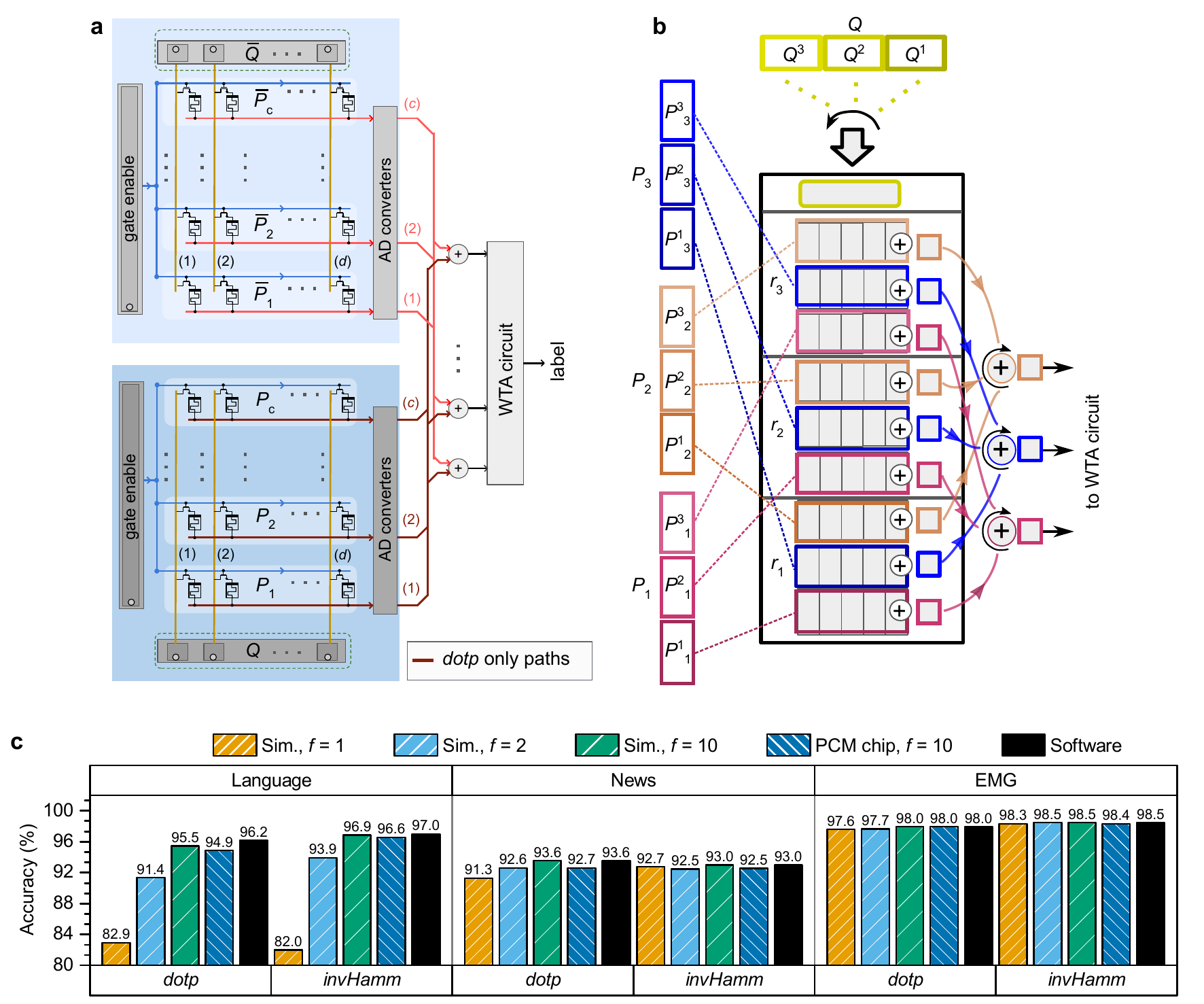}
\caption{\textbf{Associative memory search.} \textbf{a} Schematic illustration of the AM search architecture to compute the $invHamm$ similarity
metric. Two PCM crossbar arrays of $c$ rows and $d$ columns are employed. \textbf{b} Schematic illustration of the coarse grained randomization
strategy employed to counter the variations associated with the crystalline PCM state. \textbf{c} Results of the classification task show that the
experimental on-chip accuracy results compare favorably with the 10-partition simulation results and software baseline for both similarity metrics on
the three datasets.} \label{fig:2}
\end{figure}

To compute the $invHamm$ similarity metric, two memristive crossbar arrays of $c$ rows and $d$ columns are required as shown in Fig.~\ref{fig:2}\textbf{a}. The
prototype hypervectors, $P_i$, are programmed into one of the crossbar arrays as conductance states. Binary `1' components are programmed as
crystalline states and binary `0' components are programmed as amorphous states. The complementary hypervectors $\overline {P_i}$ are programmed in a
similar manner into the second crossbar array. The query hypervector $Q$ and its complement $\overline{Q}$ are applied as voltage values along the
wordlines of the respective crossbars. In accordance with the Kirchoff's current law, the total current on the $i^{th}$ bitline will be equal to the
dot-product between query hypervector and $i^{th}$ prototype hypervector. The results of this in-memory dot-product operations from the two arrays
are added in a pairwise manner using a digital adder circuitry in the periphery and are subsequently input to a winner-take-all (WTA) circuit which
outputs a `1' only on the bitline corresponding to the class of maximum similarity value. When $dotp$ similarity metric is considered, only the
crossbar encoding $P_i$ is used and the array of adders in the periphery is eliminated, resulting in reduced hardware complexity.

Experiments were performed using a prototype PCM chip to evaluate the effectiveness of the proposed implementation on three common HDC benchmarks:
language classification, news classification, and hand gesture recognition from electromyography (EMG) signals (see Methods). These tasks demand a
generic programmable architecture to support different number of inputs, classes, and data types (see Methods). In those experiments, the prototype
hypervectors (and their complements) are learned beforehand in software, and are then programmed into the PCM devices on the chip. Inference is then
performed with a software encoder and using Equation \eqref{eqn:am2} for the associative memory search, in which all multiplication operations are
performed in the analog domain (by exploiting the Ohm's law) on-chip and the remaining operations are implemented in software (see Methods and
Supplementary Note I). The software encoder was employed to precisely assess the performance and accuracy of the associative memory search alone when
implemented in hardware. The in-memory encoding scheme and its experimental validation will be presented in Sections \ref{sec:encoder} and
\ref{sec:hdsystem}.

%However, it was found that, when a naive mapping of the prototype hypervectors to the array is used, the chip-level variability (see Supplementary
%Note II) associated with the crystalline state detrimentally affects the AM search operation.

While HDC is remarkably robust to random variability and device failures, deterministic spatial variations in the conductance values
could pose a challenge. Unfortunately, in our prototype PCM chip, the conductance values associated with the crystalline state do exhibit a
deterministic spatial variation (see Supplementary Note II). However, given the holographic nature of the hypervectors, this can be easily addressed
by a random partitioning approach. We employed a coarse grained randomization strategy where the idea is to segment the prototype hypervector and to
place the resulting segments spatially distributed across the crossbar array (see Fig.~\ref{fig:2}\textbf{b}). This helps all the components of
prototype hypervectors to uniformly mitigate long range variations. The proposed strategy involves dividing the crossbar array into
\paramdef{f} equal sized partitions ($\paramdef{R}_1$,$\paramdef{R}_2$,...,$\paramdef{R}_f$) and storing a $1/f$ segment of each of the prototype
hypervectors ($\vecdef{P}_1$,$\vecdef{P}_2$,...,$\vecdef{P}_\paramdef{C}$) per partition. Here \paramdef{f} is called the \keydef{partition factor}
and it controls the granularity associated with the randomization. To match the segments of prototype hypervectors, the query vector is also split
into equal sized subvectors $\vecdef{Q}^1$,$\vecdef{Q}^2$,...,$\vecdef{Q}^f$ which are input sequentially to the wordline drivers of the crossbar.

A statistical model that captures the spatio-temporal conductivity variations was used to evaluate the effectiveness of the coarse-grained randomized
partitioning method (see Supplementary Note II). Simulations were carried out for different partition factors 1, 2 and 10 for the two similarity
metrics \scaldef{dotp} and $invHamm$ as shown in Figure~\ref{fig:2}\textbf{c}. These results indicate that the classification accuracy increases with
the number of partitions. For instance, for language classification, the accuracy improves from 82.5\% to 96\% with \scaldef{dopt} by randomizing
with a partition factor of 10 instead of 1. The experimental on-chip accuracy (performed with a partition factor of 10) is close to the 10-partition
simulation result and the software baseline for both similarity metrics on all three datasets. When the two similarity metrics are compared,
$invHamm$ provides slightly better accuracy for the same partition size, at the expense of almost doubled area and energy consumption. Therefore, for
low-power applications, a good trade-off is the use of \scaldef{dotp} similarity metric with a partition factor of 10.
%-------------------------------------------------------------------------------------------------------------------------------------------------
% ///////////////////////////////////////////////// ENCODER DESIGN ////////////////////////////////////////////////////////////////////
%-------------------------------------------------------------------------------------------------------------------------------------------------
\section{The \textit{n}-gram encoding module}
\label{sec:encoder}
\begin{figure}[h!]
\includegraphics[width=\textwidth]{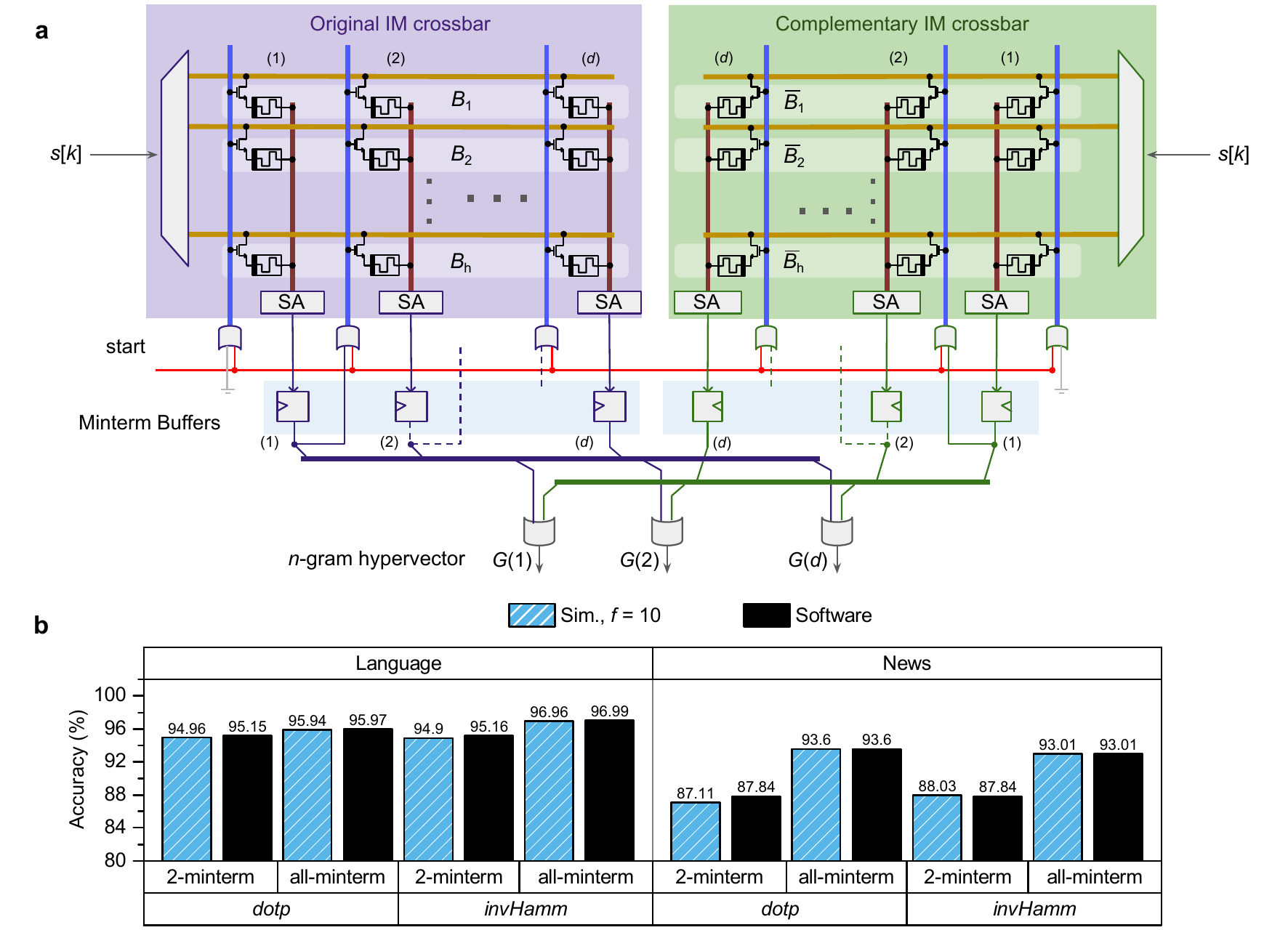}
\caption{\textbf{In-memory $n$-gram encoding based on 2-minterm.} \textbf{a} The basis hypervectors and their complements are mapped onto two crossbar arrays.
Through a sequence of in-memory logical operations the approximated $n$-gram $G$ as in Equation~(\ref{eq:encoding3}) is generated. \textbf{b}
Classification results on the language (using $n=4$) and news (using $n=5$) datasets show the performance of the 2-minterm approximation compared with the all-minterm approach.}
\label{fig:3}
\end{figure}

In this section, we will focus on the design of the $n$-gram encoding module. As described in Section \ref{sec:inmemoryhd}, one of the key operations
associated with the encoder is the calculation of the $n$-gram hypervector given by Equation (\ref{eq:encoding}). In order to find in-memory hardware
friendly operations, Equation (\ref{eq:encoding}) is re-written as the component-wise summation of $2^{n-1}$ minterms given by Equation
(\ref{eq:encoding2}).
\begin{equation} \label{eq:encoding2}
G = \bigvee_{j=0}^{2^{n-1}-1} L_{1,j}(B[1]) \wedge \rho (L_{2,j}(B[2]))\wedge \hdots \wedge \rho^{n-1} (L_{n,j}(B[n]))
\end{equation}
where the operator $L_{k,j}$ is given by
\begin{eqnarray} \label{eq:Zop}\nonumber
L_{k,j}(B[k]) &=& B[k] \text{ if } (-1)^{Z(k,j)} = 1 \\\nonumber &=& \overline{B[k]} \text{ otherwise, }
\end{eqnarray}
where $Z(k,j) = \lfloor{\frac{1}{2^k} (2j+2^{k-1})}\rfloor$, $k \in \left \{ 1,2,..., n \right \}$ is the item hypervector index within an $n$-gram and $j \in \left \{ 0,1,..., {2^{n-1}-1} \right \}$ is used to index minterms.

The representation given by Equation~(\ref{eq:encoding2}) can be mapped into memristive crossbar arrays where bitwise AND ($\wedge$) function can be realized using an in-memory read logic operation. However the number of minterms ($2^{n-1}-1$) rises exponentially with the
size $n$ of the $n$-gram, making the hardware computations costly. Therefore, it is desirable to reduce the number of minterms and to use a fixed number of minterms independent of $n$.

It can be shown that when $n$ is even, there exists a 2-minterm approximation to Equation~\ref{eq:encoding2} given by
\begin{equation} \label{eq:encoding3}
G \approx (B[1]\wedge\rho(B[2])\wedge \hdots \rho^{n-1}(B[n])) \vee (\overline{B[1]}\wedge\rho(\overline{B[2]})\wedge \hdots
\rho^{n-1}(\overline{B[n]}))
\end{equation}

We used this 2-minterm based approximation for in-memory HDC. A schematic illustration of the corresponding $n$-gram encoding system is presented in
Fig.~(\ref{fig:3}\textbf{a}). The basis hypervectors are programmed on one of the crossbars and their complement vectors are programmed on the
second. The component-wise logical AND operation between two hypervectors in Equation~(\ref{eq:encoding3}) is realized in-memory by applying one of
the hypervectors as the gate control lines of the crossbar, while selecting the wordline of the second hypervector. The result of the AND function
from the crossbar is passed through an array of sense amplifiers (SA) to convert the analog values to binary values. The binary result is then stored
in the minterm buffer, whose output is fed back as the gate controls by a single component shift to the right (left in the complementary crossbar).
This operation approximates the permutation operation in Equation~(\ref{eq:encoding3}) as a 1-bit right-shift instead of a circular 1-bit shift. By
performing these operations $n$ times, it is possible to generate the $n$-gram (the details are presented in the Methods section).

%In the first cycle, the start signal is asserted while feeding the index $s[n]$ into row decoders. At the end of the first cycle, $B[n]$ (or
%$\overline{B[n]}$) is downloaded to the minterm buffer. In subsequent cycles, a shifted version of the content of the minterm buffer is used to drive
%the gate controlling signals while choosing appropriate index into row decoder (see methods for more details). Repeating this process for $n$ cycles
%brings both the minterms on to the minterm buffer. The minterms are then merged using the AND gate array shown in Fig. \ref{fig:3}\text{a} to produce
%the approximated $n$-gram $G$ as in Equation~(\ref{eq:encoding3}).

To test the effectiveness of the encoding scheme with in-memory computing, simulations were carried out using the PCM statistical model. The training
was performed in software with the same encoding technique used thereafter for inference, and both the encoder and AM were implemented with modeled
PCM crossbars for inference.
%The proposed in-memory encoding technique can be applied to a symbols set that is mapped entirely into the IM.
%Datasets such as EMG where a spatial encoding step fuses symbols from parallel streams, leading to a large number of symbol hypervectors, cannot be entirely mapped to the IM.
The simulations were performed only on the language and news classification datasets, because for the EMG dataset the hypervectors used for the
$n$-gram encoding are generated by a spatial encoding process and cannot be mapped entirely into a fixed IM of reasonable size. From the results
presented in Fig.~\ref{fig:3}\textbf{b}, it is clear that the all-minterm approach to encoding provides the best classification accuracy in most
configurations of AM as expected. However, the 2-minterm based encoding method yields a stable and in some cases, particularly in language dataset,
similar accuracy level to that of the all-minterm approach, while significantly reducing the hardware complexity. The 2-minterm $n$-gram encoder
empirically shows to generate quasiorthogonal hypervectors closely following the behavior of the exact all-minterm encoder (see Supplementary Note
III). This approximation also appears to provide satisfactory results when $n$ is odd according to the experiments conducted (see Supplementary Note
III), even though the second minterm in Equation (\ref{eq:encoding3}) shows up in Equation (\ref{eq:encoding2}) only when $n$ is even.

%-------------------------------------------------------------------------------------------------------------------------------------------------
% ///////////////////////////////////////////////// THE COMPLETE HD computing SYSTEM  ////////////////////////////////////////////////////////////////////
%-------------------------------------------------------------------------------------------------------------------------------------------------
\section{The complete in-memory HDC system}\label{sec:hdsystem}
\begin{figure}[h!]
\includegraphics[width=\textwidth]{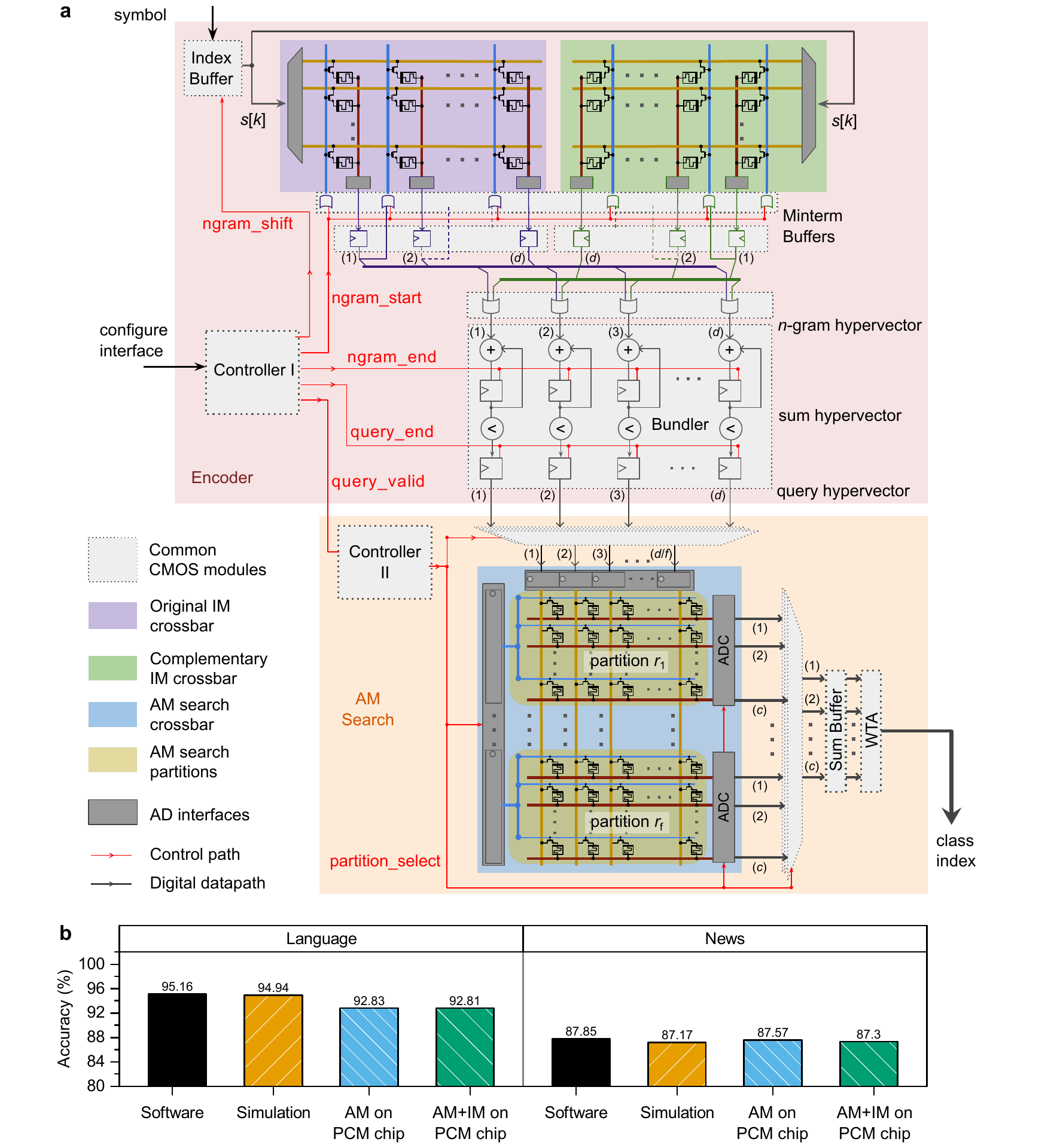}
\caption{\label{fig:4}\textbf{The complete in-memory HDC system.} \textbf{a} The schematic of architecture showing the 2-minterm encoder and
associative memory search engine employing $dotp$ metric. \textbf{b} The classification accuracy results on the news and language datasets where both
the encoding and associative memory search are performed in software, simulated using PCM model and are experimentally realized on the chip.}
\end{figure}

In this section the complete HDC system and the associated experimental results are presented. The proposed architecture comprises the
2-minterm encoder and \scaldef{dotp} similarity metric with a partition factor of 10 as this provides the best trade off between classification
accuracy and hardware complexity (see Supplementary note III). As shown in Figure~\ref{fig:4}\textbf{a}, the proposed architecture has three PCM crossbar arrays---two having $h$
rows and $d$ columns, and one having $c\cdot f$ rows and $d/f$ columns, with $f=10$.

The system includes several peripheral circuits: an index buffer, a minterm buffer, and a bundler which reside inside the encoder, whereas the AM
search module contains a sum buffer and a comparator circuit. The index buffer is located at the input of the IM to keep the indices of the symbols
in the sequence and to feed them into the crossbar rows. The bundler accumulates the $n$-gram hypervectors to produce a sum hypervector. Once
threshold is applied on the sum hypervector, the result is a prototype hypervector at the time of training or a query hypervector at the time of
inference. The controller inside the encoder module generates control signals according to the $n$-gram size and the length of the query sequence to
allow different configurations of the encoder. During inference, one segment of the query hypervector at the output buffer of the encoder is fed at a
time to the AM through an array of multiplexers so that only the corresponding partition is activated in the AM. Depending on the partition that is
selected, the relevant gates are activated through a controller sitting inside the AM search module. Finally the results in the sum buffer are sent
through a WTA circuitry to find the maximum index which provides the prediction.

\begin{table}[h!]
\caption{Performance comparison between a dedicated all-CMOS implementation and in-memory HDC with PCM crossbars \label{tab:cmos_comparision}}
\begin{tabular}{|l|r|r|r|r|r|r|}
\hline
\multirow{2}{*}{}                                & \multicolumn{3}{c|}{\textbf{All-CMOS}}                                & \multicolumn{3}{c|}{\textbf{PCM crossbar based}}                                      \\ \cline{2-7}
                                                 & \multicolumn{1}{c|}{Encoder} & \multicolumn{1}{c|}{AM search} & \multicolumn{1}{c|}{Total} & \multicolumn{1}{c|}{Encoder} & \multicolumn{1}{c|}{AM search} & \multicolumn{1}{c|}{Total} \\ \hline
\multicolumn{7}{|l|}{\textbf{Energy}}                                                                                                                                                                                                       \\ \hline
Average energy per query (nJ)                    & 1470                         & 1110                           & 2580                       & 420.8                        & 9.44                           & 430.8                      \\ \hline
\multicolumn{4}{|l|}{\textit{Improvement}}                                                                                                    & 3.50x                        & 117.5x                         & 6.01x                      \\ \hline
Exclusive modules avg. energy per query (nJ)     & 1130                         & 1100                           & 2240                       & 78.60                        & 3.30                           & 81.90                      \\ \hline
\multicolumn{4}{|l|}{\textit{Improvement}}                                                                                                    & 14.40x                       & 334.62x                        & 27.30x                      \\ \hline
\multicolumn{7}{|l|}{\textbf{Area}}                                                                                                                                                                           \\ \hline
Total area ($mm^2$)                              & 4.77                         & 2.99                           & 7.76                       & 1.39                         & 0.68                           & 2.07                       \\ \hline
\multicolumn{4}{|l|}{\textit{Improvement}}                                                                                                    & 3.43x                        & 4.38x                          & 3.74x                       \\ \hline
Exclusive modules area ($mm^2$)                  & 3.53                         & 2.38                           & 5.91                       & 0.14                         & 0.075                          & 0.21                       \\ \hline
\multicolumn{4}{|l|}{\textit{Improvement}}                                                                                                    & 24.57x                       & 31.94x                         & 27.09x   \\ \hline
\end{tabular}
\end{table}

To experimentally validate the functionality of the complete in-memory HDC architecture, we chose to implement the inference operation which
comprises both encoding (to generate the query hypervectors) and associative memory search. For faster experiments, we trained our HDC model in
software using the 2-minterm approximate encoding method described in Section \ref{sec:encoder}, that could be performed as well with our proposed
in-memory HDC architecture. This software generates the hypervectors for AM from a given dataset. Subsequently, the components of all hypervectors of
both IM and AM were programmed on individual hardware PCM devices, and the inference operation was implemented leveraging the two in-memory computing
primitives (for both 2-minterm encoding and AM search) using the prototype PCM chip (see Methods and Supplementary Note I).
%We conducted experiments on the prototype PCM chip, where all components of both IM and AM data were stored on hardware PCM devices. In the full chip
%experiment, training was performed in software, and measurements for both IM crossbars and AM crossbars at each step of the HD computing algorithm
%were taken from the PCM prototype chip for inference (see Methods).
Figure~\ref{fig:4}\textbf{b} summarizes the accuracy results with
software, the PCM statistical model, and on-chip experiment, for the language and news classification benchmarks. Compared with the previous
experiment where only AM was contained on-chip, the full chip experiment results show a similar accuracy level, indicating the minimal effect on
accuracy when porting the IM into PCM devices with in-memory $n$-gram encoding. Furthermore, the accuracy level reported in this experiment is
close to the accuracy reported with the software for the same parametric configuration of the HD inference model.

Finally, to benchmark the performance of the system in terms of energy consumption, the digital submodules in the system-level architecture (marked
with dotted boundaries in Figure~\ref{fig:4}) that fall outside the PCM crossbars arrays were synthesized using \unit[65]{nm} CMOS technology. The synthesis
results of these modules can be combined with the performance characteristics of PCM crossbar arrays to arrive at figures of merit such as energy,
area and throughput of the full system (see Methods). Furthermore, PCM crossbar sections were implemented in CMOS distributed standard cell registers with
associated multiplier adder tree logic and binding logic respectively for AM and IM to construct a
complete CMOS HD processor with the intention of comparing against the figures of merits of the PCM crossbar based architecture proposed in this paper.
%Due to limitations in EDA tools used for synthesizing the HD processor, dimensionality $d$ had to be limited to 2000. The results obtained from the tools were then
%appropriately scaled up to derive the results for dimensionality $d$=10000 processor to have a fair comparison with PCM based HD inference system.
A comparison of the performance between all-CMOS approach versus the PCM crossbar based approach is presented in Table~\ref{tab:cmos_comparision}. As seen
in the table, a $6.01\times$ improvement in total energy efficiency and $3.74\times$ reduction in area is obtained with the introduction of the PCM crossbar modules.
The encoder's energy expense for processing a query reduces by a factor of $3.50$ with the PCM crossbar implementation whereas that of the AM search
module reduces by a factor of $117.5$. However, these
efficiency factors are partially masked by the CMOS peripheral circuitry that is common to both implementations, specifically that in the encoder
module which accounts for the majority of its energy consumption. When peripheral circuits are ignored and only the parts of the design that are
exclusive to each approach are directly compared to each other, $14.4\times$ and $334\times$ energy savings and $24.5\times$ and $31.9 \times$ area savings are obtained for the encoder and AM search
module, respectively. It remains part of the future work to investigate methods in which peripheral modules are designed more energy efficiently so
that the overall system efficiency can be improved further.

\section{Conclusion}
Hyperdimensional computing is a brain-inspired computational framework that is particularly well-suited for the emerging computational paradigm of
in-memory computing. We presented a complete in-memory HDC system whose two main components are an encoder and an associative memory search engine.
The main computations are performed in-memory with logical and dot product operations on memristive devices. Due to the inherent robustness of HDC to
errors, it was possible to approximate the mathematical operations associated with HDC to make it suitable for hardware implementation, and to use
analog in-memory computing without significantly degrading the output accuracy. Our architecture is programmable to support different hypervector
representations, dimensionality, number of input symbols and of output classes to accommodate a variety of applications. Hardware/software
experiments using a prototype PCM chip delivered accuracies comparable to software baselines on language and news classification benchmarks with
$10,000$-dimensional hypervectors, making this work the largest experimental demonstration of HDC with memristive hardware to date. It is also
arguably one of the largest and most significant experimental demonstrations of in-memory logic, a field of study initiated by the seminal work of
Borghetti et al.\cite{Y2010borghettiNature}. These experiments used hardware PCM devices to implement both in-memory encoding and associative memory
search, thus demonstrating the hardware functionality of all the operations involved in a generic HDC processor for learning and inference.  A
comparative study performed against a system-level design implemented using \unit[65]{nm} CMOS technology showed that the in-memory HDC approach
could result in $>6\times$ end-to-end savings in energy. By designing more energy-efficient peripheral circuits and with the potential of scaling PCM
devices to nanoscale dimensions \cite{Y2011xiongScience}, these gains could increase several fold. The in-memory HDC concept is also applicable to
other types of memristive devices based on ionic drift \cite{Y2010waserNT} and magnetoresistance \cite{Y2015kentNatureNanotechnology}. Future work
will be focused on taking in-memory HDC beyond learning and classification to perform advanced cognitive tasks alongside with data compression and
retrieval on dense storage devices as well as building more power efficient peripheral hardware to harness the best of in-memory computing.

\section*{Author contributions}
All authors collectively conceived the idea of in-memory hyperdimensional computing. G.K. performed the experiments and analyzed the results under
the supervision of M.L.G, A.R., and A.S. G.K., M.L.G., A.R., and A.S. wrote the manuscript with input from all authors.

\section*{Acknowledgments}
This work was supported in part by the European Research Council through the European Union's Horizon 2020 Research and Innovation Program under
Grant 682675 and in part by the European Union's Horizon 2020 Research and Innovation Program through the project MNEMOSENE under Grant 780215. We
would like to thank Evangelos Eleftheriou for managerial support.

\section{Competing financial interests}
The authors declare no competing financial interests.
\bibliographystyle{naturemag}
\bibliography{library}

\clearpage
\section*{Methods}
\subsection*{PCM-based hardware platform}
The experimental hardware platform is built around a prototype phase-change memory (PCM) chip that contains PCM cells that are
based on doped-Ge$_2$Sb$_2$Te$_2$ (d-GST) and are integrated into the prototype chip in \unit[90]{nm} CMOS baseline technology. In addition to the
PCM cells, the prototype chip integrates the circuitry for cell addressing, on-chip ADC for cell readout, and voltage- or current-mode cell
programming. The experimental platform comprises the following main units:
\begin{itemize}
    \item a high-performance analog-front-end (AFE) board that contains the digital-to-analog converters (DACs) along with discrete electronics, such as power supplies, voltage, and current reference sources,
    \item an FPGA board that implements the data acquisition and the digital logic to
    interface with the PCM device under test and with all the electronics of the AFE board, and
    \item a second FPGA board with an embedded processor and Ethernet connection that implements the overall system control and data management as well as the interface with the host computer.
\end{itemize}

The prototype chip~\cite{Y2010closeIEDM} contains 3 million PCM cells, and the CMOS circuitry to address, program and readout any of these 3 million
cells. In the PCM devices used for experimentation, two \unit[240]{nm}-wide access transistors are used in parallel per PCM element (cell size is
\unit[50]{F$^2$}). The PCM array is organized as a matrix of 512 word lines (WL) and 2048 bit lines (BL). The PCM cells were integrated into the chip
in 90 nm CMOS technology using the key-hole process ~\cite{Y2007breitwischVLSI}. The bottom electrode has a radius of $\sim 20$~nm and a length of
$\sim 65$~nm. The phase change material is $\sim 100$ nm thick and extends to the top electrode, whose radius is $\sim 100$ nm. The selection of one
PCM cell is done by serially addressing a WL and a BL. The addresses are decoded and they then drive the WL driver and the BL multiplexer. The single
selected cell can be programmed by forcing a current through the BL with a voltage-controlled current source. It can also be read by an 8-bit on-chip
ADC. For reading a PCM cell, the selected BL is biased to a constant voltage of \unit[300]{mV} by a voltage regulator via a voltage $V_\mathrm{read}$
generated via an off-chip DAC. The sensed current, $I_\mathrm{read}$, is integrated by a capacitor, and the resulting voltage is then digitized by
the on-chip 8-bit cyclic ADC. The total time of one read is \unit[$1$]{$\mu$s}. For programming a PCM cell, a voltage $V_\mathrm{prog}$ generated
off-chip is converted on-chip into a programming current, $I_{\mathrm{prog}}$. This current is then mirrored into the selected BL for the desired
duration of the programming pulse. The pulse used to program the PCM to the amorphous state (RESET) is a box-type rectangular pulse with duration of
\unit[400]{ns} and amplitude of \unit[450]{$\mu$A}. The pulse used to program the PCM to the crystalline state (SET) is a ramp-down pulse with total
duration of approximately \unit[12]{$\mu$s}. The access-device gate voltage (WL voltage) is kept high at \unit[2.75]{V} during the programming
pulses. These programming conditions were optimized in order to have the highest on/off ratio and to minimize device-to-device variability for binary
storage.

\subsection*{Datasets to evaluate in-memory HDC}
We target three highly relevant learning and classification tasks to evaluate the proposed in-memory HDC architecture. These tasks demand a generic
programmable architecture to support different number of inputs, classes, and data types as shown in Table \ref{tab:dataset_conf}. In the following,
we describe these tasks that are used to benchmark the performance of in-memory HDC in terms of classification accuracy.
\begin{enumerate}
\item \textbf{Language classification}:  In this task, HDC is applied to classify raw text composed of Latin characters into their respective
language. The training texts are taken from the Wortschatz Corpora~\cite{language_trainset} where large numbers of sentences (about a million bytes
of text) are available for 22 European languages. Another independent dataset, Europarl Parallel Corpus~\cite{language_testset}, with 1,000 sentences
per language is used as the test dataset for the classification. The former database is used for training 22 prototype hypervectors for each of the
languages while the latter is used to run inference on the trained HDC model. For the subsequent simulations and experiments with the language
dataset we use dimensionality
\paramdef{D}$=10,000$ and $n$-gram size
\paramdef{N}$=4$.

We use an item memory (IM) of 27 symbols, representing the 26 letters of the Latin alphabet plus whitespace character. Training is performed using
the entire training dataset, containing a labeled text of $120,000$--$240,000$ words per language. For inference, a query is composed of a single
sentence of the test dataset, hence in total 1,000 queries per language are used.

\item \textbf{News classification}: The news dataset comprises a
database of Reuters news articles, subjected to a light weight pre-processing step, covering 8 different news genres~\cite{NewsDataset}. The
pre-processing step removes frequent ``stop'' words and words with less than 3 letters \cite{HD_news}. The training set has 5400+ documents while the
testing set contains 2100+ documents. For the subsequent simulations and experiments with news dataset we use dimensionality
\paramdef{D}$=10,000$ and $n$-gram size \paramdef{N}$=5$.
Similar to the language task, we use an IM of 27 symbols, representing the 26 letters of the Latin alphabet plus whitespace character. Training is
performed using the entire training dataset, where all labeled documents pertaining to the same class are merged into a single text. This merged text
contains $8,000$--$200,000$ words per class. For inference, a query is composed of a single document of the test dataset.

\item \textbf{Hand gesture recognition from Electromyography (EMG) signals}:

In this task, we focus on use of HDC in a smart prosthetic application, namely hand gesture recognition from a stream of EMG signals. A
database~\cite{emgdataset} that provides EMG samples recorded from four channels covering the forearm muscles is used for this benchmark. Each
channel data is quantized into 22 intensity levels of electric potential.
The sampling frequency of the EMG signal is \unit[500]{Hz}.

A label is provided for each time sample. The label varies from 1 to 5 corresponding to five classes of performed gestures. This dataset is used to
train an HDC model to detect hand gestures of a single subject. For training on EMG dataset, a spatial encoding scheme is first employed to fuse data
from the four channels so the IM has four discrete symbols, and it is paired with a \emph{continuous} item memory to jointly map the 22 intensity
levels per channel (the details on encoding procedure for EMG dataset are explained in Supplementary Note IV). The pairing of IM and CIM allows a
combination of orthogonal mapping with distance proportionality mapping. The spatial encoding creates one hypervector per time sample.

Then, a temporal encoding step is performed, whereby $n$ consecutive spatially encoded hypervectors are combined into an $n$-gram. For the subsequent
simulations and experiments with EMG dataset we use dimensionality \paramdef{D}$=10,000$ and $n$-gram size \paramdef{N}$=5$. Training and inference
are performed using the same EMG channel signals from the same subject, but on non-overlapping sections of recording. The recording used for training
contains 1280 time samples after down-sampling by a factor of 175. For inference, 780 queries are generated from the rest of recording, where each
query contains 5 time samples captured with the same down-sampling factor.
\end{enumerate}

\begin{table}[]
\centering
\begin{threeparttable}
\begin{tabular}{|c||c|c|c|c|c|c|c|}
\hline \multirow{2}{*}{Dataset} & \multirow{2}{*}{Input type} & \multirow{2}{*}{$n$-gram size} & \multirow{2}{*}{\# of channels} & \multicolumn{2}{c|}{Item Memory (IM)} & \multicolumn{2}{c|}{Associative Memory (AM)} \\ \cline{5-8}
                         & & & & \# Symbols $h$       & Dimensionality $d$       & Dimensionality $d$           & \# Classes $c$          \\ \hline
Language                 & Categorical & 4 & 1 & 27            & 10,000   & 10,000    & 22         \\ \hline
News                     & Categorical & 5 & 1 & 27 & 10,000   & 10,000    & 8          \\ \hline
EMG                      & Numerical   & 5 & 4 & 4             & 10,000                & 10,000 & 5          \\ \hline
\end{tabular}
\end{threeparttable}
\caption{Architecture configurations and hyperparameters used for the tree different tasks}
 \label{tab:dataset_conf}
\end{table}
For the different tasks, Table \ref{tab:dataset_conf} provides details on the desired hypervector representations, and different hyperparameters
including the dimension of hypervectors, the alphabet size, the $n$-gram size, and the number of classes.
%Table \ref{tab:dataset_conf} provides details on the dimensions of the IM and AM for the different tasks.
%
For EMG dataset, the hypervectors for the encoding operation are drawn by binding items from a pair of IM and continuous IM (Supplementary Note IV).
In hardware implementation of in-memory HDC, the IM and AM may be distributed into multiple narrower crossbars in case electrical/physical limitations arise.

\subsection*{Coarse grained randomization}
The programming methodology followed to achieve the coarse grained randomized partitioning in memristive crossbar for AM search is explained in
the following steps.
First, we split all prototype hypervectors ($\vecdef{P}_1$,$\vecdef{P}_2$,...,$\vecdef{P}_c$) into $\paramdef{f}$ subvectors of equal length where $\paramdef{f}$ is the partition factor.
For example, subvectors from the prototype hypervector of the first class are denoted as:
($\vecdef{P}_1^1$,$\vecdef{P}_1^2$,...,$\vecdef{P}_1^f$).
Then the crossbar array is divided into $\paramdef{f}$ equal sized partitions
($\paramdef{R}_1$,$\paramdef{R}_2$,...,$\paramdef{R}_f$).
Each partition must contain $\paramdef{D}/{\paramdef{f}}$ rows and $\paramdef{C}$ columns.
A random permutation \paramdef{E} of numbers 1 to \paramdef{C} is then selected.
Next, the first subvector from each class ($\vecdef{P}_1^1$,$\vecdef{P}_2^1$,...,$\vecdef{P}_c^1$) is programmed into the first partition $\paramdef{R}_1$
such that each subvector fits to a column in the crossbar partition.
%A logical 1 component in the hypervector is programmed with the high conductance state whereas logical 0 component is programmed with the low conductance state of the memristor.
The order of programming of subvectors into the columns in the partition is determined by the
previously selected random permutation \paramdef{E}.
The above steps must be repeated to program all the remaining partitions ($\paramdef{R}_2$,$\paramdef{R}_3$,...,$\paramdef{R}_f$).

The methodology followed in feeding query vectors during inference is detailed in the following steps.
%\begin{enumerate}
First, we split query hypervector \vecdef{Q} into $f$ subvectors $\vecdef{Q}^1$,$\vecdef{Q}^2$,...,$\vecdef{Q}^f$ of equal length.
Then, we translate $\vecdef{Q}^i$ component values into voltage levels and apply onto the wordline drivers in
the crossbar array. Bitlines corresponding to the partition $\paramdef{R}_i$ are enabled. Depending on the belonging class, the partial dot products
are then collected onto respective destination in sum buffer through the A/D converters at the end of $\paramdef{R}_i$ partition of the array.
The above procedure is repeated for each partition $r_i$.
Class-wise partial dot products are accumulated together in each iteration and updated in the sum buffer. After the \paramdef{f}-th iteration, full dot
product values are ready in the sum buffer. The results are then compared against each other using a WTA circuit to find the maximum value to assign
its index as the predicted class.
%\end{enumerate}

\subsection*{Experiments on associative memory search}
In order to obtain the prototype hypervectors used for AM search, training with HDC is first performed in software on the three datasets described in
the section ``Datasets to evaluate in-memory HDC''. For the language and news datasets, XOR-based encoding (see Section \ref{sec:inmemoryhd}) is used
with $n$-gram size of $n=4$ and $n=5$, respectively. For the EMG dataset, an initial spatial encoding step creates one hypervector per time sample.
Then, a temporal encoding step is performed, whereby $n$ consecutive spatially encoded hypervectors are combined into an $n$-gram with XOR-based
encoding and $n=5$. The detailed encoding procedure for EMG dataset is explained in Supplementary Note IV.

Once training is performed, the prototype hypervectors are programmed on the prototype PCM chip.
In the experiment conducted with $invHamm$ as the
similarity metric, $d\times c \times 2$ devices on the PCM prototype chip are allocated. Each device in the first half of the address range (from 1 to $d\times c$) is programmed with a component of a prototype hypervector $P_i$, where $i=1,...,c$. Devices in the second half of the array are programmed with components of the complementary prototype
hypervectors. The exact programming order is determined by the partition factor ($f$) employed in the coarse grained randomized partitioning scheme. For $f=10$ used in the experiment, devices from first address up to $1000 \times c$-th address are programmed with content of the first partition, i.e., the first segment of
each of the prototype hypervector. The second set of $1000 \times c$ addresses is programmed with content of the second partition and so on. As the hypervector
components are binary, devices mapped to the logical 1 components and devices mapped to logical 0 components are programmed to the maximum (approximately \unit[20]{$\mu$S})
and minimum conductance (approximately \unit[0]{$\mu$S}) levels respectively. The devices are programmed in a single-shot (no iterative program-and-verify algorithm is used) with a single RESET/SET pulse for minimum/maximum conductance devices.

Once the programming phase is completed, the queries from the testing set of a given task are encoded. Only for the experiments of Section
\ref{sec:am}, the query hypervectors are generated using the same software HD encoder used for training. In the experiments of Section
\ref{sec:hdsystem}, the query hypervectors are generated with in-memory encoding using the prototype PCM chip as described in the section
``Experiments on the complete in-memory HDC system''.

The associative memory search on a given query hypervector is performed using the prototype PCM chip as follows. The components of the query
hypervector carrying a value 1 trigger a read (\unit[300]{mV} applied voltage) on the devices storing the corresponding components of prototype
hypervectors, thus realizing the analog multiplications through Ohm's law of the in-memory dot-product operation. The same procedure is performed
with the complementary query hypervector on the devices storing complementary prototype hypervectors. The resulting current values are digitized via
the on-chip ADC, transferred to the host computer and class-wise summed up in software according to the predetermined partition order to obtain
class-wise similarity values (see Supplementary Note I). The class with the highest similarity is assigned as the predicted class for the given
query. For experiments with $dotp$ as the similarity metric, the devices attributed to complementary prototype hypervectors are not read when forming
the class-wise aggregate.

\subsection*{More details on the 2-minterm encoder}
In order to generate a $n$-gram hypervector in $n$ cycles, the crossbar is operated using the following procedure.
During the first cycle, $n$-gram encoding is initiated by asserting the \keydef{start} signal while choosing the index of $n$-th symbol $\scaldef{s}[\paramdef{N}]$.
    This enables all the gate lines in both crossbar arrays and the wordline corresponding to $\scaldef{s}[\paramdef{N}]$ to be activated.
    The current released onto the bitlines passed through the sense amplifiers should ideally match the logic levels of $\vecdef{B}[{\paramdef{N}]}$ in first array
    and $\overline{\vecdef{B}[{\paramdef{N}}]}$ in the second array.
    The two \textit{'minterm buffers'} downstream of the sense amplifier arrays register the two hypervectors by the end of the first cycle.
    During subsequent $j$-th ($1<\scaldef{j}\leq\paramdef{N}$) cycles, the gate lines are driven by the right shifted version of the incumbent values on the minterm buffers---effectively implementing permutation---while row decoders are fed with symbol $s[\paramdef{N}-\scaldef{j}+1]$; the left shift is used for the second crossbar.
    This ensures that the output currents on the bitlines correspond to the component-wise logical AND between the permuted minterm buffer values and the next basis hypervector $\vecdef{B}[{\paramdef{N}-\scaldef{j}}]$ (complement for the second array).
The expression for the value stored on the left-side minterm buffers at the end of $\scaldef{j}$-th cycle is given by
$ \prod_{k=1}^{\scaldef{j}} \rho^{\scaldef{j}-k}\,\vecdef{B}[{\paramdef{N}-k+1}] $.
The product of the complementary hypervectors $ \prod_{k=1}^{\scaldef{j}} \rho^{\scaldef{j}-k}\,\overline{\vecdef{B}[{\paramdef{N}-k+1}]} $ is stored in the right-side minterm buffers.
At the end of the $\paramdef{N}$-th cycle, the two minterms are available in the minterm buffers.
The elements in the minterm buffers are passed onto the OR gate array following the minterm buffers (shown in Figure~\ref{fig:3}), such that inputs to the array have matching indices from the two minterm vectors.
At this point, the output of the OR gate array reflects the desired $n$-gram hypervector from 2-minterm $n$-gram encoding.

After $n$-gram encoding, the generated $n$-grams are accumulated and binarized. In the hardware implementation, this step is realized inside the bundler module shown in Figure~\ref{fig:4}.
The threshold applied to binarize the sum hypervector components is given by: $$ l \cdot \left(\frac{1}{2^{n-log(k)}}\right) $$
where $l$ is the length of the sequence, $n$ is the $n$-gram size, and $k$ is the number of minterms used for the binding operation in the encoder.

\subsection*{Experiments on the complete in-memory HDC system}
For the experiments concerning the complete in-memory HDC system, training with HDC is first performed in software on the language and news datasets. 2-minterm encoding (Equation \eqref{eq:encoding3}) is used with $n$-gram size of $n=4$ and $n=5$, respectively.

After training is performed, $h\times d\times 2$ devices are allocated on the PCM chip for storing IM and
the complementary IM in addition to $d\times c$ devices allocated for AM. The IM and complementary IM hypervectors are programmed on PCM devices in a single-shot with RESET/SET pulses for logical 0/1 components. The prototype hypervectors of the AM are programmed as described in the section ``Experiments on associative memory search'', with the exception that the complementary prototype
hypervectors are not programmed since $dotp$ is used as the similarity metric.

During inference, for every query to be encoded, the IM and complementary IM are read from the prototype PCM chip. In-memory read logic (AND) is
performed by thresholding the read current values from the on-chip ADC in software to emulate the sense amplifiers of the eventual proposed hardware
at each step of the 2-minterm $n$-gram encoding process (see Supplementary Note I). The other operations involved in the encoder that are not
supported by the prototype PCM chip such as the 1-bit right-shift permutation, storing of the intermediate results in the minterm buffers, ORing the
results of the original and complementary minterm buffers, and the bundling of $n$-gram hypervectors, are implemented in software. Once the encoding
of the query hypervector is completed, the associative memory search is carried out on that query hypervector as specified in the section
``Experiments on associative memory search'' with $dotp$ as the similarity metric.

\subsection*{Performance, energy estimation and comparison}
In order to evaluate and benchmark energy efficiency of the proposed architecture, a cycle-accurate register transfer level (RTL) model of a complete
CMOS design that has equivalent throughput to that of the proposed in-memory HDC system architecture is developed (see Supplementary Note V).
A testbench infrastructure is then built to verify the correct behavior of the model.
Once the behavior is verified, the RTL model is synthesized in UMC \unit[65]{nm} technology node using Synopsys Design Compiler.
Due to limitations in EDA tools used for synthesizing the CMOS-based HDC, dimensionality $d$ had to be limited to $2,000$.
The post-synthesis netlist is then verified using the same stimulus vectors applied during behavioral simulation.
During post-synthesis netlist simulation, the design is clocked at \unit[440]{MHz} frequency to create a switching activity file in value change dump (VCD) format for inference of 100 language classification queries.
Then, the energy estimation for the CMOS modules is performed by converting average power values reported by Synopsys Primetime which takes the netlist and the activity file from the previous steps as the inputs.
A typical operating condition with voltage \unit[1.2]{V} and temperature \unit[25]{C} is set as the corner for the energy estimation of the CMOS system.
Further energy and area results were obtained for $d$ values 100, 500, 1000 in addition to 2000. Then the results were extrapolated to derive the energy and area estimates for dimensionality $d=10,000$ to have a fair comparison with in-memory HDC system.

The energy/area of the proposed in-memory HDC system architecture is obtained by adding the energy/area of the modules that are common with the full CMOS design described above,
together with the energy of PCM crossbars and the analog/digital peripheral circuits exclusive to the in-memory HDC architecture.
Parameters based on the prototype PCM chip in \unit[90]{nm} technology used in the experiments are taken as the basis for the PCM-exclusive energy/area estimation.
The parameters of the sense amplifiers which are not present in the PCM hardware platform but present in the proposed in-memory HD encoder are taken from the \unit[65]{nm} current latched sense amplifier presented by Chandoke et al.~\cite{sense_amp}.
Parameters used for PCM exclusive energy estimation are shown in Table \ref{tab:pcm_energy}.
\begin{table}[]
\caption{Parameters for PCM-exclusive energy estimation \label{tab:pcm_energy}}
\begin{tabular}{|l|r|r|}
\hline
\multicolumn{3}{|l|}{\textbf{Common parameters}}                                                         \\ \hline
\textbf{Parameter}            & \multicolumn{2}{l|}{\textbf{Value}}                                      \\ \hline
Read voltage                  & \multicolumn{2}{r|}{\unit[0.1]{V}}                                               \\ \hline
Current on conducting devices & \multicolumn{2}{r|}{\unit[1]{$\mu$A}}                                                 \\ \hline
Unit device area              & \multicolumn{2}{r|}{\unit[0.2]{$\mu$m$^2$}}                            \\ \hline
\multicolumn{3}{|l|}{\textbf{Module-specific parameters}}                                                \\ \hline
\textbf{Parameter}             & \multicolumn{1}{l|}{\textbf{Encoder}} & \multicolumn{1}{l|}{\textbf{AM}} \\ \hline
Readout time                 & \unit[2.8]{ns}                                 & \unit[100]{ns}                            \\ \hline
Active devices per query      & 145,000                               & 66,000                           \\ \hline
Energy per sense amp read      & \unit[9.8]{fJ}                                  & \multicolumn{1}{c|}{-}           \\ \hline
Energy per ADC read           & \multicolumn{1}{c|}{-}                & \unit[12]{pJ}                             \\ \hline
\end{tabular}
\end{table}

\end{document}